\documentclass[aps,onecolumn,twoside,secnumarabic,12pt,balancelastpage,amsmath,amssymb,nofootinbib,hyperref=pdftex]{revtex4}

\usepackage{color}         
\usepackage{multirow}
\usepackage{graphics}      
\usepackage[pdftex]{graphicx}      
\usepackage{longtable}     
\usepackage[english]{babel}
\usepackage{amsmath}
\usepackage{epsf}          
\usepackage{bm}            
\usepackage{verbatim}			
\usepackage{cancel}
\usepackage[colorlinks=true]{hyperref}  

\begin{document}

\begin{flushright}
\today
\end{flushright}

\vspace{0.07in}

\noindent
\begin{center}

{\bf\large Discovering Axion-Like Particles with Photon Fusion at the ILC}

\vspace{0.5cm}
{Noah Steinberg}

{\it Leinweber Center for Theoretical Physics \\
Physics Department, University of Michigan \\
Ann Arbor, MI 48109-1040 USA}\\
\end{center}

\noindent
{\it Abstract:}
Experimental searches for Axion-Like Particles (ALPs) which couple to the electroweak bosons span over a wide range of ALP masses, from MeV searches at beam-dump experiments, to TeV searches at the LHC. Here we examine an interesting range of parameter space in which the ALP couples only to hypercharge. In the GeV to hundreds of GeV mass range, the contribution of an ALP to light by light scattering can be significant. By making simple kinematic cuts, we show that the ILC running at $\sqrt{s} = 250\,\rm{GeV}$ or $\sqrt{s} = 500\,\rm{GeV}$ can discover ALPs in this range of masses with significantly smaller couplings to the SM than previous experiments, down to  $g_{aBB} = 10^{-3}\,\rm{TeV}^{-1}$.
\vfill\eject

\section {Introduction}
Models of Axion-Like Particles (ALPs) with masses spanning the sub eV to TeV scale have been extensively explored in recent years\cite{Ringwald:2014vqa}. This is due to the genericness of ALPs; they are simply pseudo-scalars that couple to the SM gauge bosons and possibly fermions and are singlets with respect to the SM gauge symmetries\cite{Beacham:2019nyx, weinberg}. Additionally they may contribute to the DM content of the Universe\cite{Duffy:2009ig}. Obviously this definition covers a wide landscape of theories, which is why attempts to narrow down this landscape begin with investigating ALPs coupling to a single SM species, e.g.\ only to fermions, or to gluons, etc.\ One of the simplest incarnations of these models is giving the ALPs a coupling to hypercharge and nothing else (i.e.\ assuming all other couplings subdominant). The Lagrangian for this theory is simply:
\begin{equation}
\mathcal{L} = \mathcal{L_{SM}} + \frac{1}{2}(\partial_{\mu}a)^{2} - \frac{1}{2}m_{a}^{2}a^{2} - \frac{g_{aBB}}{4}a\tilde{B}^{\mu\nu}B_{\mu\nu},
\end{equation}
where $a$ is the Axion-Like Particle and ($\tilde{B}^{\mu\nu}$)$B^{\mu\nu}$ is the (dual)field strength tensor of $U(1)_{\rm{Y}}$ hypercharge. 
\vskip 0.12in
This dimension 5 operator, suppressed by the coupling $g_{aBB}\,(\rm{GeV}^{-1})$, after electroweak symmetry breaking induces couplings of the ALP to the $Z$ boson and to photons. These couplings allow one to search for ALPs in a myriad of processes including rare $Z$ and Higgs decays and other electroweak processes\cite{Florez:2021zoo, Steinberg:2021iay, dEnterria:2021ljz}. Especially interesting is the coupling of the ALP to two photons via the effective operator $\mathcal{O}_{a\tilde{F}F} \propto a\tilde{F}_{\mu\nu}F^{\mu\nu}$. This operator induces an effective $a-\gamma-\gamma$ vertex which for example allows for associated production of ALPs with a single photon (mono-photon) as well as provides a decay channel of the ALP into two photons. This is the dominant decay mode in this model even for $m_{a} \gg m_{Z}$ because the decay width to $\gamma\gamma$, $\gamma Z$, and $ZZ$ are proportional to $c_{W}^{4}$, $c_{W}^{2}s_{W}^{2}$, and $s_{W}^{4}$ respectively. The restricted phase space for decays involving a $Z$ decreases those branching fractions for $m_{a}$ near and below threshold. 
	As the production and decay of an ALP in this model is a purely electroweak process, we will show that the ILC, a next generation $e^{+} e^{-}$ collider, is the ideal laboratory to discover and study ALPs. 
\vskip 0.12in
\section{ALP Production and constraints}
\begin{figure}[t]
\begin{center}
\includegraphics[width=14cm]{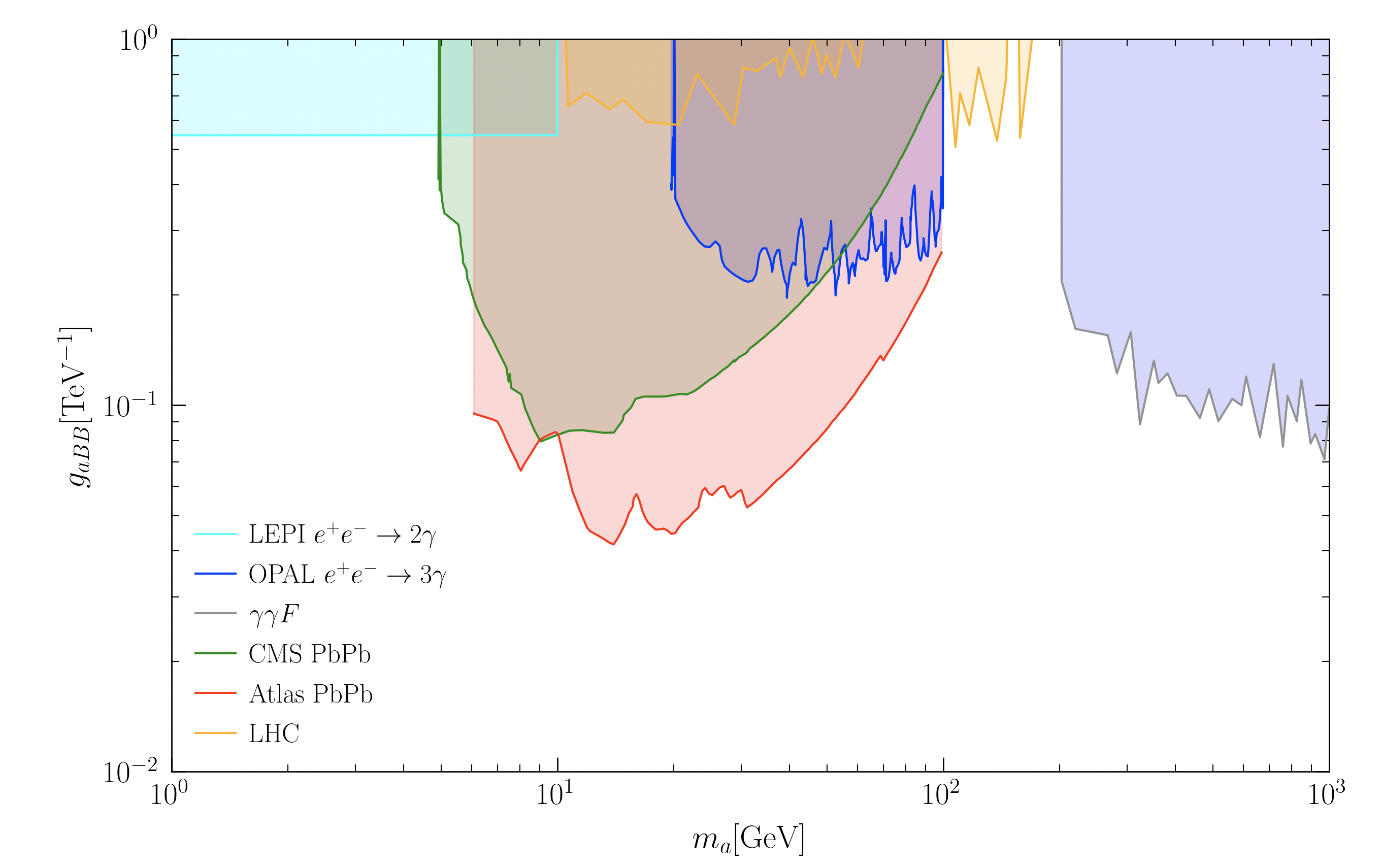}
\caption{Summary of constraints on ALPs coupled to hypercharge. Figure adapted from\cite{Jaeckel:2015jla}}
\label{fig:former_constraints}
\end{center}
\end{figure}

The most promising search channel for heavy ($m_{a} > 100\,\rm{MeV})$ ALPs coupled to hypercharge depends strongly on the collider type. In \cite{Knapen:2016moh,Aad:2020cje,Sirunyan:2018fhl}, ALPs with couplings to photons were searched for in Ultra-Peripheral Lead-Lead ion collisions at CMS and ATLAS. Here the process of interest is $\rm{PbPb}\rightarrow \gamma\gamma\rm{PbPb}$ where quasi-real photons from two incoming lead ions scatter from each other at large impact parameter, leaving the lead ions intact. These analyses take advantage of the $Z^{4}$ enhancement in coherent photon-photon luminosity, enhancing the discovery sensitivity for low mass (5 - 100 GeV) scalars and pseudo-scalars which couple to photons, even with the reduced PbPb luminosity (only $1\,\rm{nb}^{-1}/\rm{year})$. The coherent enhancement becomes suppressed past $\gamma\gamma$ invariant masses of $\approx$ 200 GeV (as nuclear breakup becomes more probable), limiting the mass reach of this method. Similar analyses are possible in $p-p$ collisions ($pp\rightarrow p\gamma\gamma p$), though the intact protons must be forward tagged to overcome the large hadronic backgrounds at the LHC\cite{Baldenegro:2018hng}. 
\vskip 0.12in
Other non-exclusive $\gamma\gamma$ final states can been used to search for ALPs, such as in vector boson fusion\cite{Florez:2021zoo} where the back to back jets from the partonic process as well as the two final state photons are used for triggering. This gives competitive constraints in the 10 MeV to 100 GeV range with ALP couplings down to $0.5\,\rm{TeV}^{-1}$ . Finally in the high mass region $200\,\rm{GeV} < m_{a} < 2.6\,\rm{TeV}$, ATLAS searches for spin 0 resonances in diphoton final states\cite{Aaboud:2017yyg} have been re-interpreted by employing photon distribution functions in the proton to produce the most stringent constraints at these high masses\cite{Bauer:2018uxu}. 
\vskip 0.12in
A summary of these constraints on $g_{aBB}$ is given in Fig.~\ref{fig:former_constraints}.
As one can see, couplings at the $10^{-1}\,\rm{TeV}^{-1}$ level are currently being probed depending on the mass range of interest, with the strongest constraints between 10 and 100 GeV, and above 200 GeV, with a curious gap in sensitivity in the 100 - 200 GeV region. In this region (and at even smaller values of $m_{a}$) we will show that the ILC running at either 250 GeV or 500 GeV will increase the sensitivity to this model between 10 and 200 GeV, and to even higher masses at ILC500. In the next section we briefly review the capabilities of the ILC.
\section{ILC}
The International Linear Collider (ILC) is a proposed next generation $e^{+}e^{-}$ collider\cite{Bambade:2019fyw}. The ILC hopes to study the Standard Model (SM) with unprecedented precision at $\sqrt{s} = m_{Z},\,250\,\rm{GeV},\,500\,\rm{GeV}$ and possibly 1 TeV, with a particularly close eye on precision electroweak and Higgs physics. Besides testing the SM, the ILC also has unique capabilities to search for Beyond the SM (BSM) physics. Currently the proposed run plan for the ILC is to collect $2\,\rm{ab}^{-1}$ at 250 GeV and $3\,\rm{ab}^{-1}$ at 500 GeV center of mass. Obviously these integrated luminosities are preliminary but we would like to stress that the reach of this specific BSM model is highly dependent on the integrated luminosity collected at each center of mass energy. Combining these large luminosities with a clean collision environment mostly free of the strong interactions which challenge the LHC make the ILC an excellent laboratory to search for rare, new physics. Additionally, the ILC's highly granular detectors allow for excellent photon identification and isolation, allowing photons with very small separations from each other to be identified, which is crucial for searching for low mass particles which decay to pairs of photons. Information on the two main proposals for ILC detectors, the ILD and the SiD can be found in~\cite{ILD,SiD}.
\section{Equivalent Photon Approximation (EPA)}
\begin{figure}[t]
\begin{center}
\includegraphics[width=10cm]{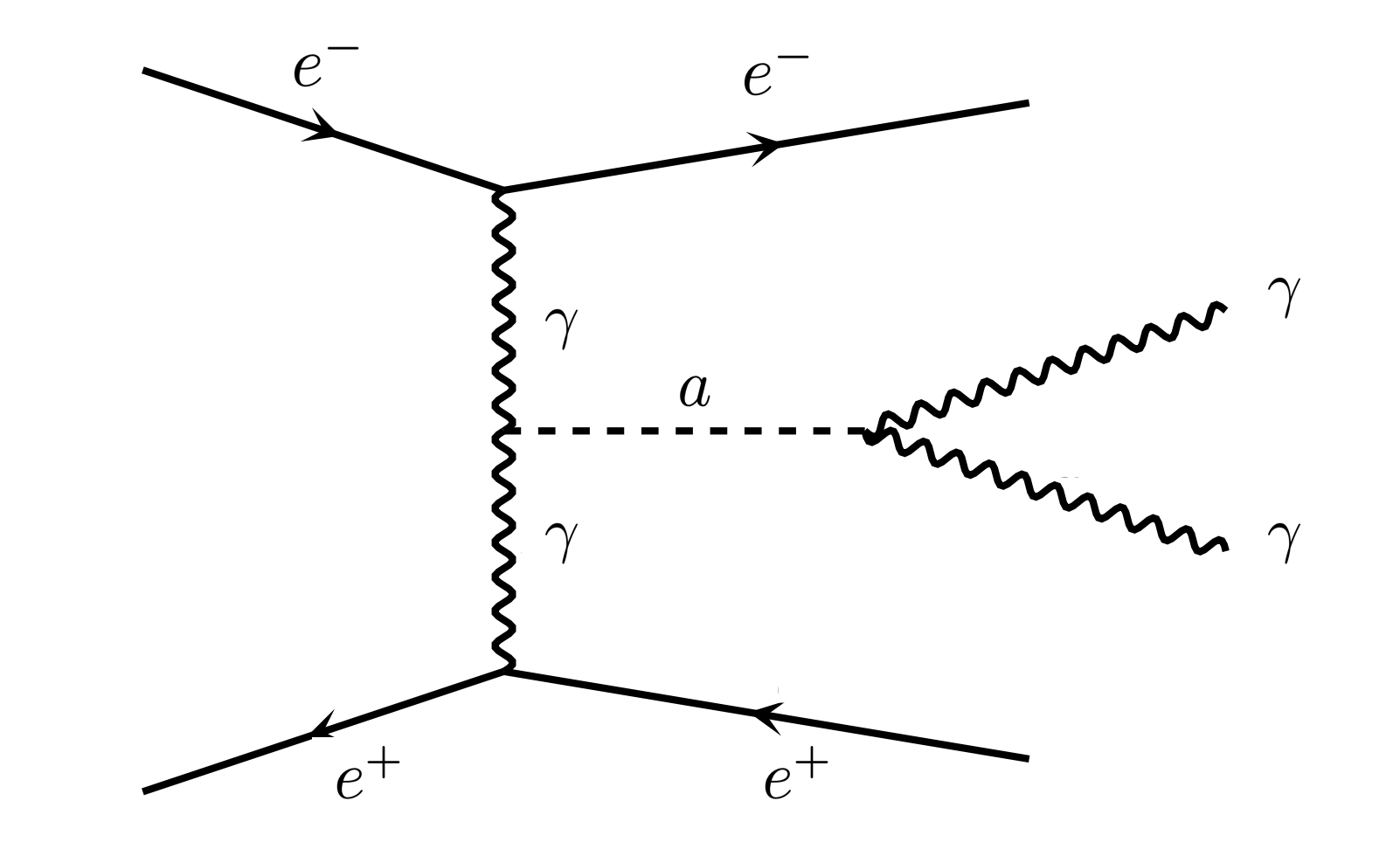}
\caption{ALP production via photon fusion from EPA photons and subsequent decay into a pair of photons. Only $s$ channel contribution is shown. Generated with $\rm{MadGraph5\_aMC@NLO}$.}
\label{fig:ALPfusion_diagram}
\end{center}
\end{figure}

The ILC, in addition to being an $e^{+}e^{-}$ collider, is also a $\gamma\gamma$ collider, by virtue of the equivalent photon approximation\cite{Budnev:1974de}. In the EPA, photons with small virtuality are emitted via bremstrahlung almost collinearely with an incoming lepton beam, such that $Q^{2} = -q^{2} \approx 0$, and the photons can be treated as real. Cross sections are computed by using these almost real photons as the incoming particles in the hard scattering process, and then convolving this cross section with the photon luminosity function. 
\vskip 0.12in
In more detail, the cross section for the production of a state $X$ is given in the EPA by
\begin{equation}\label{eq:1}
\sigma(e^{+} e^{-} \rightarrow  e^{+} e^{-} X)(s)= \int ds_{\gamma\gamma}\frac{d\mathcal{L}(s_{\gamma\gamma})}{d s_{\gamma\gamma}}\sigma(\gamma\gamma \rightarrow X, s_{\gamma\gamma}),
\end{equation}
where, $\sqrt{s}$ is the center of mass energy of the lepton beams and $\sqrt{s_{\gamma\gamma}}$ is the invariant mass of the interacting photon pair. The approximation is accurate up to corrections of order $Q^{2}/s_{\gamma\gamma}$ where $Q^{2} = - q^{2}$ is the virtuality of the radiated photons, so the approximation accuracy increases as we consider larger values of $s_{\gamma\gamma}$. In the EPA, the photon luminosity function $\mathcal{L}(s_{\gamma\gamma})$ can be computed as
\begin{equation}\label{eq:2}
\frac{d\mathcal{L}(s_{\gamma\gamma})}{d s_{\gamma\gamma}} = \frac{1}{s}\int \frac{N(x_{1})}{x_{1}} N(x_{2} = s_{\gamma\gamma}/x_{1}s)dx_{1}.
\end{equation}
Here $N(x)$ is the Weizsacker-Williams photon spectrum which gives the distribution of photons emitted as a function of the energy fraction, $x = (E_{e} - E^{'}_{e})/E_{e} = E_{\gamma}/E_{e}$, and $x_{1}(x_{2})$ is the fraction of the $e^{-}(e^{+})$ energy carried by the incoming photon.
\begin{equation}\label{eq:3}
N(x) = \frac{\alpha}{2\pi}\left( (1 + (1 - x)^{2})\text{log}\left( \frac{Q^{2}_{\rm{max}}}{Q^{2}_{\rm{min}}}\right) - 2m^{2}_{e}x^{2}(\frac{1}{Q^{2}_{\rm{min}}} - \frac{1}{Q^{2}_{\rm{max}}}) \right).
\end{equation}
\vskip 0.12in
In Eq.~\ref{eq:2}, $Q^{2}_{max(min)}$ is the maximum (minimum) virtuality of the EPA photons. We take $Q^{2}_{min} = m_{e}^{2}x^{2}/(1 - x)$ and $Q^{2}_{max} = 2\,\rm{GeV^{2}}$. In the center of momentum frame of the $\gamma\gamma$ system, $s_{\gamma\gamma} = x_{1}x_{2}s$. This relationship fixes $x_{2}$ as a function of $s_{\gamma\gamma}$ and $x_{1}$. 
\begin{figure}[t]
\begin{center}
\includegraphics[width=14cm]{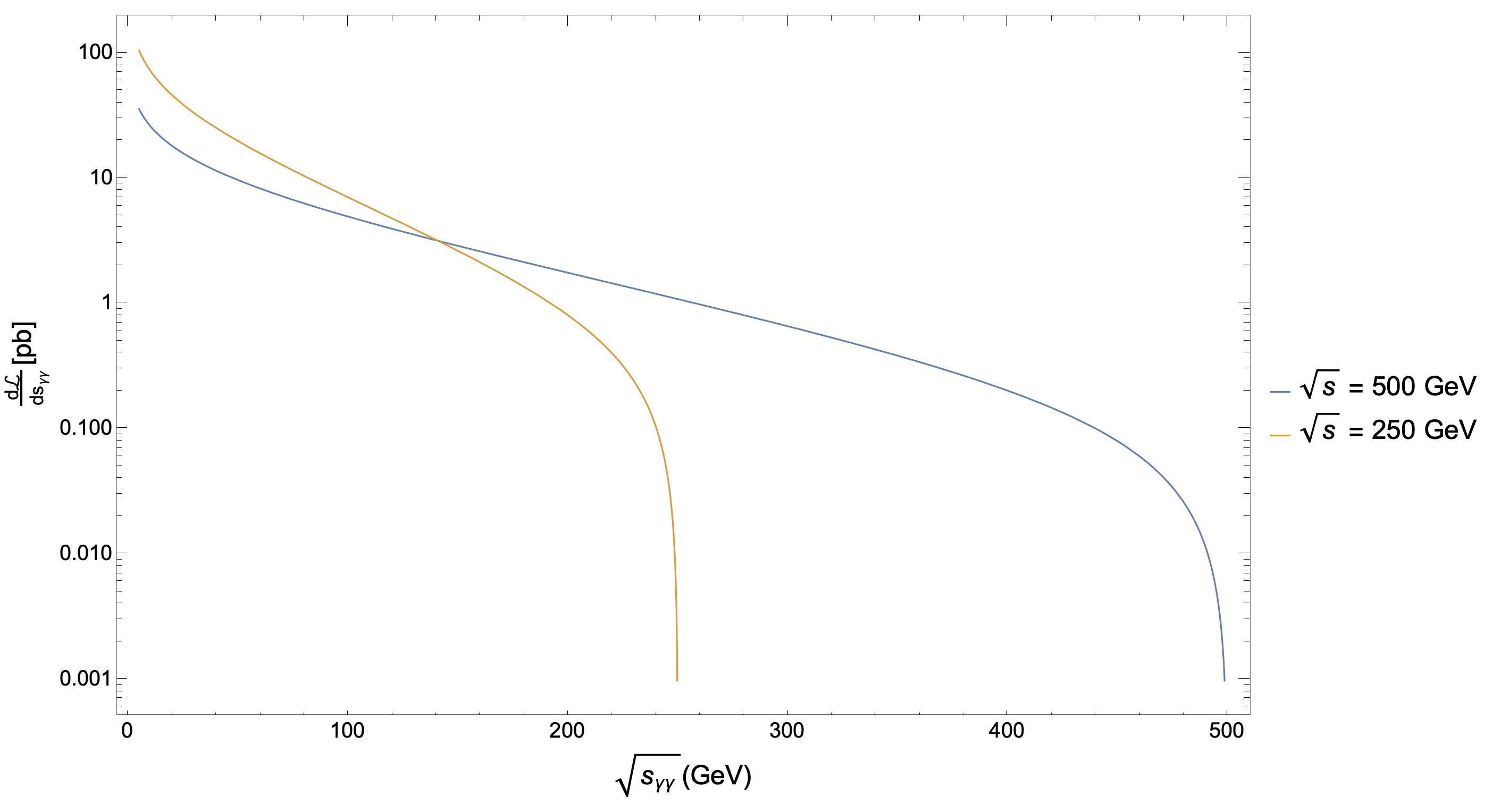}
\caption{Photon Luminosity, $\frac{d\mathcal{L}}{d \sqrt{s_{\gamma\gamma}}}$, for both ILC250 and ILC500. The $\gamma\gamma$ luminosity steeply falls as $s_{\gamma\gamma}$ increases, dropping to 0 when $s_{\gamma\gamma}$ approaches  $s$. Note the logarithmic y axis.}
\label{fig:luminosity}
\end{center}
\end{figure}
\vskip 0.12in
We use $\rm{MadGraph5\_aMC@NLO}$\cite{Alwall:2014hca} to simulate this process at the ILC by selecting the hard $\gamma\gamma\rightarrow \gamma\gamma$ process and choosing the "photons from electrons" beam mode. This selects EPA photons as the incoming particles from a lepton beam with center of mass energy, $\sqrt{s}$, chosen by the user. This process is pictured in Fig.~\ref{fig:ALPfusion_diagram}.
\vskip 0.12in
It is interesting to note that even though the ILC nominal run plan includes significant beam time with polarized leptons, this has little effect on the EPA cross section. The effect of the polarization of an incoming lepton on the polarization of an outgoing EPA photon is of order $x = E_{\gamma}/E_{l}$, so that only for $x\approx 1$ are the photons fully polarized. For small values of $x$ the photon is effectively unpolarized\cite{Philipsen:1992gz}.
To validate MadGraph's implementation of the EPA, we have verified that the numerical predictions for the production of $e^{+}e^{-}$ pairs via photon fusion match our analytical calculations for the same process. This ensures the accuracy of MadGraph's calculation of the photon luminosity function. 
\vskip 0.12in
The photon luminosity for both ILC250 and ILC500 is shown in Fig.~\ref{fig:luminosity}. Note the steep decrease in luminosity as $s_{\gamma\gamma}$ increases, with the photon luminosity for ILC250 dropping to 0 at $\sqrt{s_{\gamma\gamma}} = 250\,\rm{GeV}$. Events are showered using Pythia8\cite{Sjostrand:2014zea} and run through a fast detector simulation implementation of a generic ILC detector using Delphes\cite{deFavereau:2013fsa,Zarnecki}
\section{Backgrounds and Signal Selection}
\begin{figure}[t]
\begin{center}
\includegraphics[width=13cm]{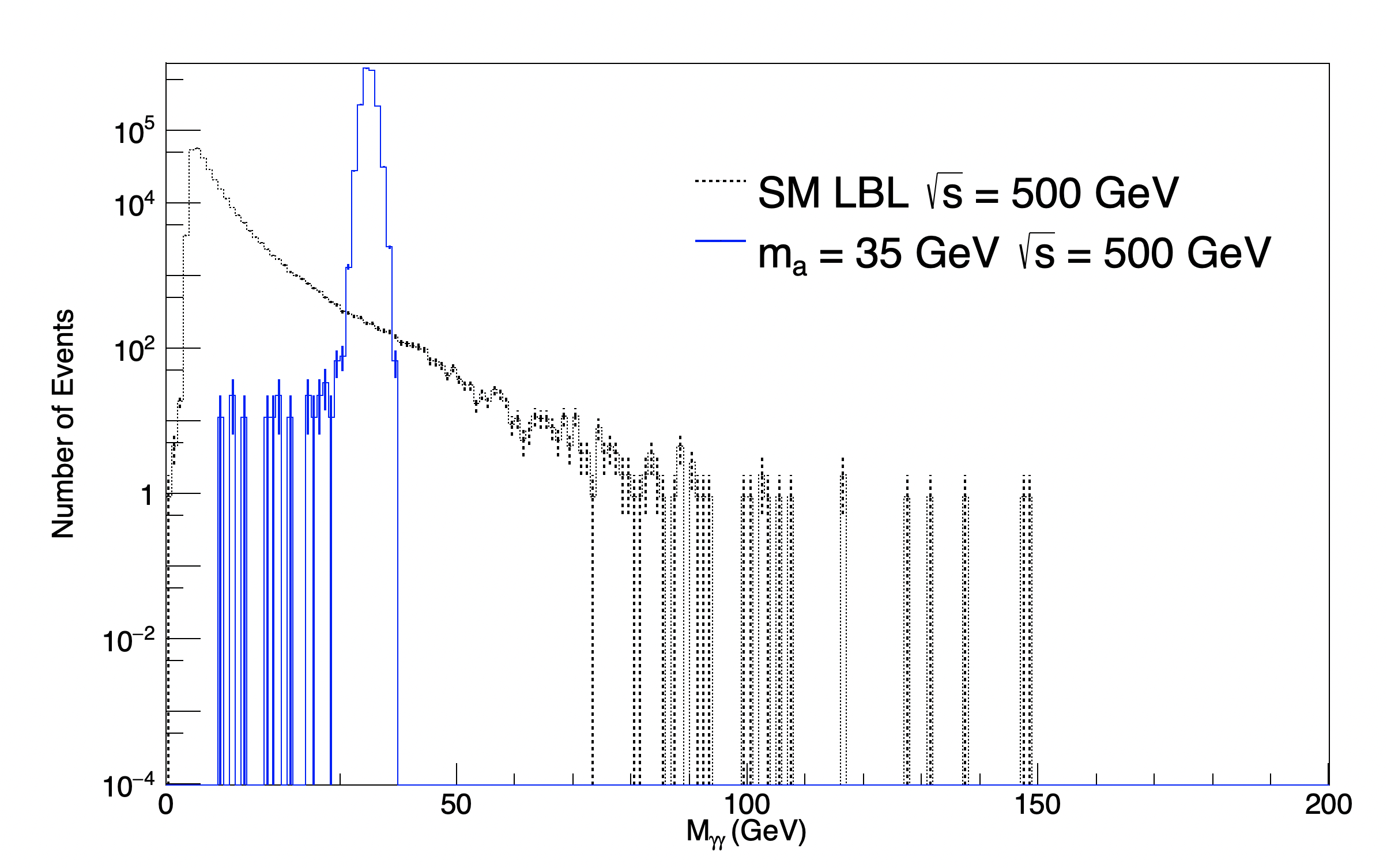}
\caption{Invariant mass distribution of the $\gamma\gamma$ system at ILC with $\sqrt{s} = 500$ GeV. Shown in blue solid line is the distribution from an ALP with $m_{a} = 35$ GeV and $g_{aBB} = 10^{-3}$ GeV, and the SM LBL distribution in dotted black.}
\label{fig:minv}
\end{center}
\end{figure}

\begin{figure}[t]
\begin{center}
\includegraphics[width=13cm]{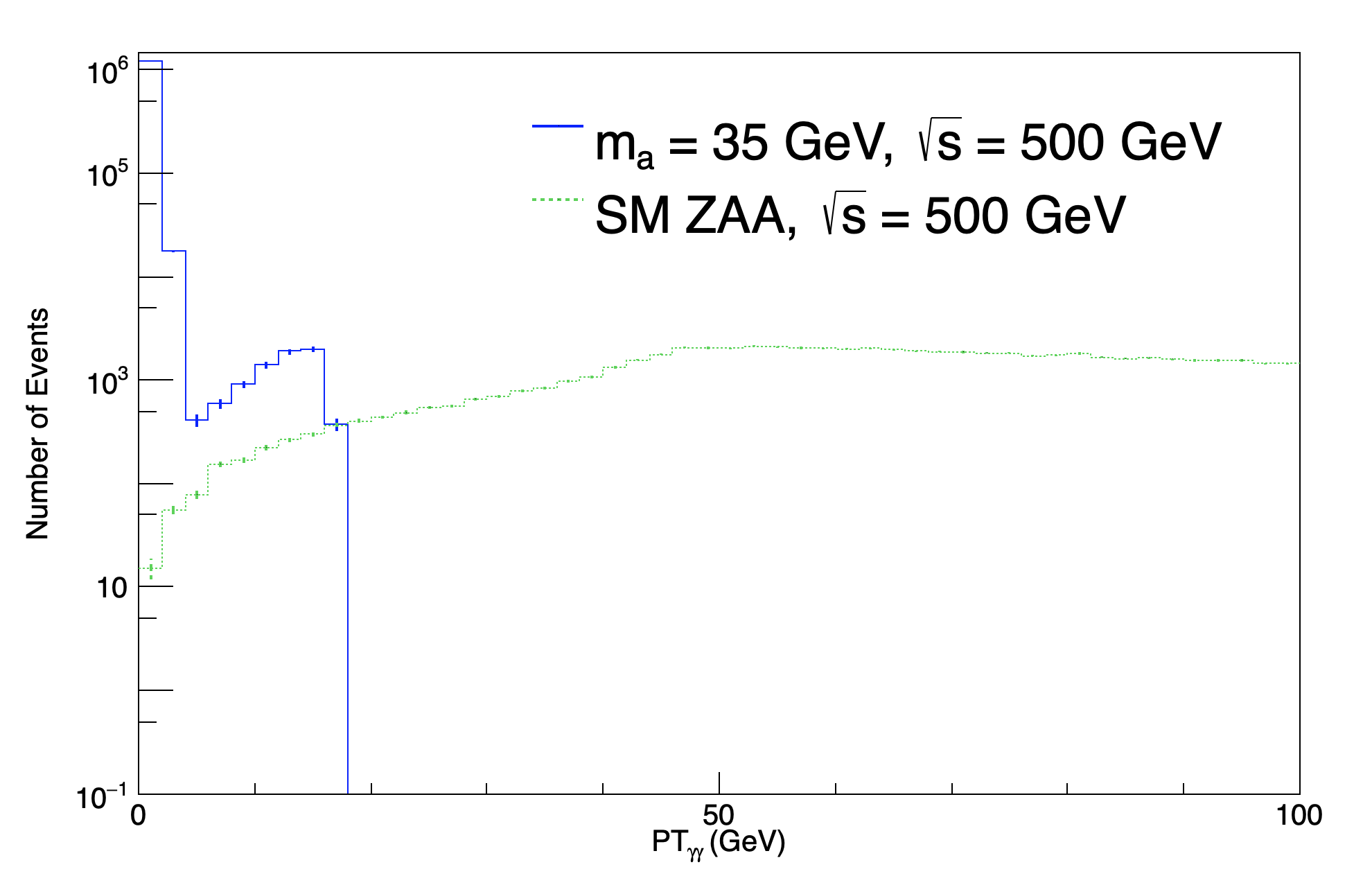}
\caption{$\text{PT}_{\gamma\gamma}$ distribution at ILC with $\sqrt{s} = 500$ GeV. Shown in blue solid line is the distribution from an ALP with $m_{a} = 35$ GeV and $g_{aBB} = 10^{-3}$ GeV. The $Z\gamma\gamma$ distribution is the dotted green line.}
\label{fig:PTaa}
\end{center}
\end{figure}

The signal selection criteria for this model is two isolated photons with no other activity and the recoiling electrons undetected. For our signal model several benchmark ALP masses were chosen based on running at $\sqrt{s} = 250$ GeV or $\sqrt{s} = 500$ GeV, as larger invariant photon masses can be reached in the later stage. For $\sqrt{s} = 250$ GeV ALP masses are chosen from 5 - 150 GeV in steps of 2 GeV, and for $\sqrt{s} = 500$ GeV ALP masses from 5 - 350 GeV were chosen with the same step size. 
\vskip 0.12in
For simulations we chose an ALP-hypercharge coupling, $g_{aBB}$, with value $10^{-3}\,\rm{GeV^{-1}}$. In searching for this signal, several background processes must be suppressed. The first and most obvious background is ordinary SM light by light (LBL) scattering, which was first observed by ATLAS in ultra peripheral PbPb collisions\cite{LBL}. We simulated this process at 1-loop level in MadGraph using the $\rm{sm\_loop\_qed\_qcd\_Gmu}$ model file with all default parameters, generating 400,000 events\cite{Hirschi:2015iia}. This process can be most efficiently suppressed via an invariant mass cut on the two final state photons, as the SM LBL $m_{\gamma\gamma}$ peaks at small values and exponentially falls off at higher values of $m_{\gamma\gamma}$. This can be seen in Fig.~\ref{fig:minv}, where we plot the invariant mass distribution of the final state photons for SM LBL scattering and our ALP signal, with $m_{a} = 35$ GeV. 
\vskip 0.12in
Another interesting background is $e^{+}e^{-} \rightarrow Z\gamma\gamma \rightarrow \bar{\nu}_{l}\nu_{l}\gamma\gamma$ with the neutrinos escaping undetected. This background can be suppressed via a cut on the transverse momentum of photon system, $\text{PT}_{\gamma\gamma} < 5$ GeV, which is equivalent to a cut on missing transverse energy, $\cancel{\it{E}}_{T}$. As can be seen in Fig.~\ref{fig:PTaa}, the $Z\gamma\gamma$ background can be efficiently suppressed with very little effect on the ALP signal. Additionally we require that the pseudorapidity of each final state photon satisfies $\eta_{\gamma} < 2.4$.
With these selection cuts, the ILC running at $\sqrt{s} = 500\,\rm{GeV}$($\sqrt{s} = 250\,\rm{GeV}$) with an integraed luminosity of $2\,\rm{ab}^{-1}$ could produce as many as $240$($160$) ALPs with a mass of $35\,\rm{GeV}$ with coupling $g_{aBB} = 10^{-4}\,\rm{GeV}^{-1} = 10^{-1}\,\rm{TeV}^{-1}$.
\section{Discovery reach of the ILC}
\begin{figure}[t]
\begin{center}
\includegraphics[width=13cm]{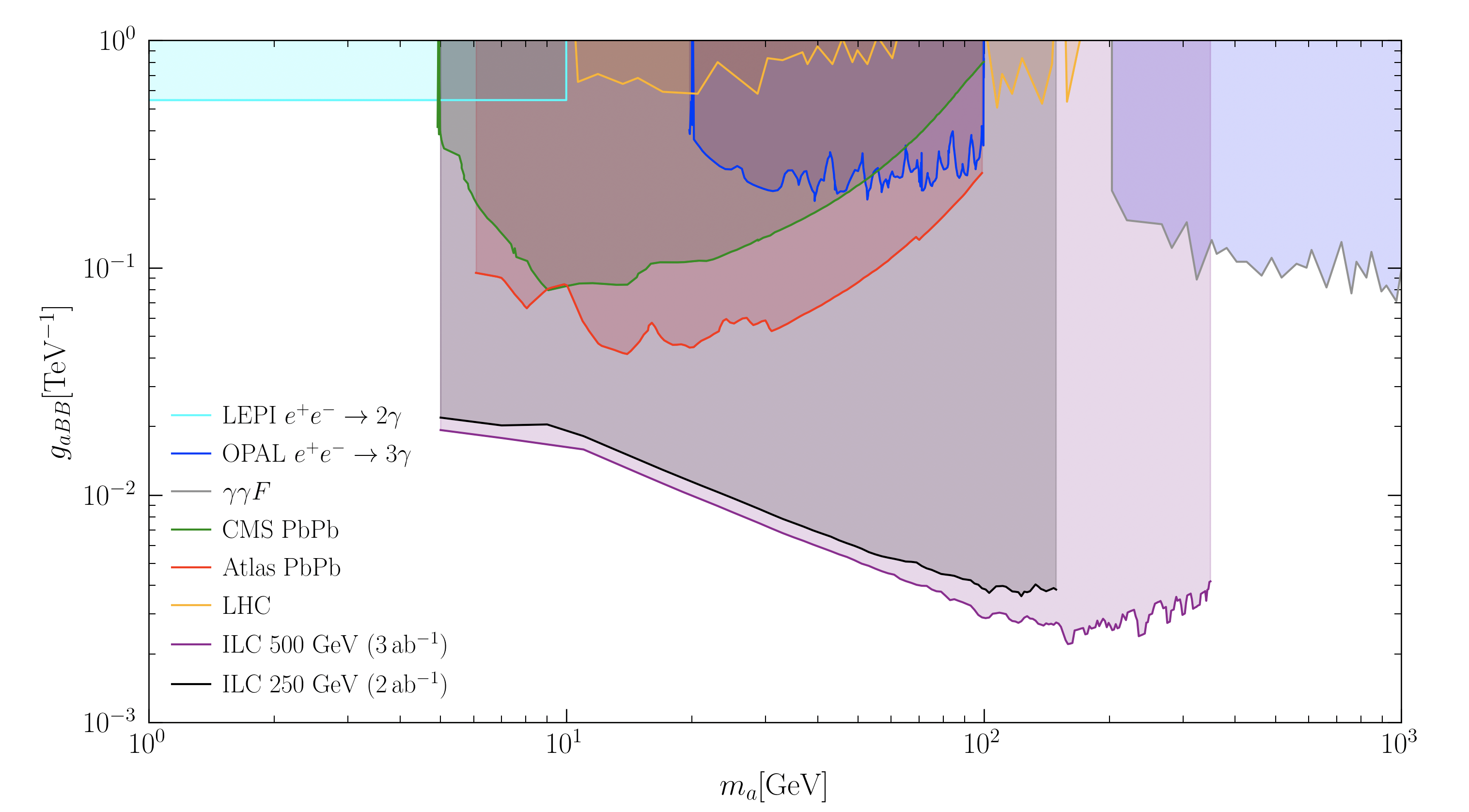}
\caption{ILC500 (purple) and ILC250 (black) reach as a function of $m_{a}$. The ILC250 will significantly open up new discovery territory in the 50 - 150 GeV region with the ILC500 improving this capability up to masses of 350 GeV.}
\label{fig:constraints}
\end{center}
\end{figure}

To compute the reach of the ILC in the $m_{a},g_{aBB}$ plane, we utilize the exact Asimov significance, $Z^{A}$, for exclusion\cite{Bhattiprolu:2020mwi}. Consider signal and background processes with poisson means $s$ and $b$ respectively. The $p$-value for exclusion of the signal model if $n$ events are observed is given by 
\begin{equation}
p_{\rm{excl}}(n,b,s) = \sum_{k=0}^{n}P(k | s+b) = Q(n+1, s+b) = \frac{\Gamma(n+1, s+b)}{\Gamma(n+1)},
\end{equation}
where $Q(a,x) = \Gamma(a, x)/\Gamma(a)$ is the regularized upper incomplete gamma function~\cite{math_gam}. In computing the exact Asimov significance the exact $p$-value for exclusion is used with the number of events $n$ replaced by its expected mean, which for exclusion is simply the mean number of background events $b$. This leads to the following expression for the exclusion $p$-value
\begin{equation}
p_{\rm excl}^{\rm Asimov} = Q(b+1,s+b).
\end{equation}
The exact Asimov significance computed in this way is a more conservative estimate of the exclusion significance and does not suffer from the counter-intuitive flaws of the median significance as noted in~\cite{Bhattiprolu:2020mwi}.
\vskip 0.12in
At each ALP mass we compute the expected number of background events, $b$, given the cuts discussed above. We then invert $Q(b+1,s+b) = 0.05$ to compute the number of signal events which gives us our 95\% confidence limit on the ALP coupling. Note that this is always well defined because the upper incomplete gamma function is monotonic. Finally, we translate the number of signal events into an expected upper limit on $g_{aBB}$. Note that in extracting limits on $g_{aBB}$ we use the narrow width approximation where
\begin{equation}
\sigma(\gamma\gamma \rightarrow a \rightarrow \gamma\gamma) \propto \sigma(\gamma\gamma \rightarrow a ) \times \rm{BR}( a\rightarrow\gamma\gamma ) \propto g_{aBB}^{2},
\end{equation}
which is valid for all masses and couplings considered here.
To compare against similar existing experimental bounds, we plot the projected upper limits on $g_{aBB}$ as a function of ALP mass against bounds from other past and current experiments in Fig.~\ref{fig:constraints}.
\vskip 0.12in
At first glance it is easy to see that running this search at $\sqrt{s} = 500\,\rm{GeV}$ gives access to much higher ALP masses due to the larger energy available from each incoming lepton. At ILC500, ALPs with masses of almost 350 GeV can be produced while having a feeble coupling to hypercharge of almost $10^{-3}\,\rm{TeV}^{-1}$. In contrast ILC250 is limited to ALP masses below 150 GeV with similar values of $g_{aBB}$. It is interesting to note that nowhere in this $m_{a},g_{aBB}$ region does ILC250 outperform ILC500. This is due to a complicated interplay between the ALP production cross section as well as the photon flux which is itself a function of the center of mass energy of the lepton beams. Suffice to say that ILC500 has significantly better discovery capabilities for ALPs in this mass range. 
\section{Conclusion}
We have shown that photon-photon scattering mediated by an ALP can be a powerful probe of Axion-Like Particles with couplings to hypercharge. Using simple kinematic cuts as well as taking advantage of the relatively background free nature of this signal leads to a powerful search strategy for these particles. Additionally, this improves the physics case for the ILC and other $e^{+}e^{-}$ colliders, showing that the clean and highly controlled collider environment provides excellent access to weakly coupled new physics. 

\section{Acknowledgements}
We thank Advanced Research Computing at the University of Michigan, Ann Arbor for their computational resources. This work was supported by the DOE under grant DE-SC0007859. N. Steinberg is supported by a fellowship from the Leinweber Center for Theoretical Physics.

\newpage{\pagestyle{empty}\cleardoublepage}

\end{document}